\documentclass{svjour3}                     
\smartqed  
\usepackage{graphicx}
\usepackage[square]{natbib}

\newcommand{\dertot}[2]{\frac{d #1}{d #2}}

\begin{document}

\title{Light-time computations for the BepiColombo radioscience
experiment}



\author{G. Tommei        \and
        A. Milani \and D. Vokrouhlick\'{y}
}


\institute{G. Tommei, A. Milani \at Department of Mathematics, University of
                     Pisa, Largo B. Pontecorvo 5, 56127 Pisa, Italy\\
           \email{tommei@dm.unipi.it}
                     \and
           A. Milani \at
           \email{milani@dm.unipi.it}
                      \and
           D. Vokrouhlick\'{y} \at Institute of Astronomy, Charles
                     University, V Hole\v{s}ovi\v{c}k\'ach 2, CZ-18000
                     Prague 8, Czech Republic \\
           \email{vokrouhl@cesnet.cz}
             }

\date{Received: date / Accepted: date}

\maketitle

\begin{abstract}
The radioscience experiment is one of the on board experiment of the
Mercury ESA mission BepiColombo that will be launched in 2014. The
goals of the experiment are to determine the gravity field of Mercury
and its rotation state, to determine the orbit of Mercury, to
constrain the possible theories of gravitation (for example by
determining the post-Newtonian (PN) parameters), to provide the
spacecraft position for geodesy experiments and to contribute to
planetary ephemerides improvement. This is possible thanks to a new
technology which allows to reach great accuracies in the observables
range and range rate; it is well known that a similar level of
accuracy requires studying a suitable model taking into account
numerous relativistic effects. In this paper we deal with the
modelling of the space-time coordinate transformations needed for the
light-time computations and the numerical methods adopted to avoid
rounding-off errors in such computations.

\end{abstract}

\section{Introduction}
\label{intro}

BepiColombo is an European Space Agency mission to be launched in
2014, with the goal of an in-depth exploration of the planet Mercury;
it has been identified as one of the most challenging long-term
planetary projects. Only two NASA missions had Mercury as target in
the past, the Mariner 10, which flew by three times in 1974-5 and
Messenger, which carried out its flybys on January and October 2008,
September 2009 before it starts its year-long orbiter phase in March
2011.

The BepiColombo mission is composed by two spacecraft to be put in
orbit around Mercury. The radioscience experiment is one of the on
board experiments, which would coordinate a gravimetry, a rotation and
a relativity experiment, using a very accurate range and range rate
tracking. These measurements will be performed by a full 5-way link
(\cite{iess01}) to the Mercury orbiter; by exploiting the frequency
dependence of the refraction index, the differences between the
Doppler measurements (done in Ka and X band) and the delay give
information on the plasma content along the radiowave path. In this
way most of the measurements errors introduced can be removed,
improving of about two orders of magnitude with respect to the past
technologies. The accuracies that can be achieved are $10$ cm in range
and $3 \times 10^{-4}$ cm/s in range rate.

\begin{figure}[h]
\begin{center}
\includegraphics[width=6cm]{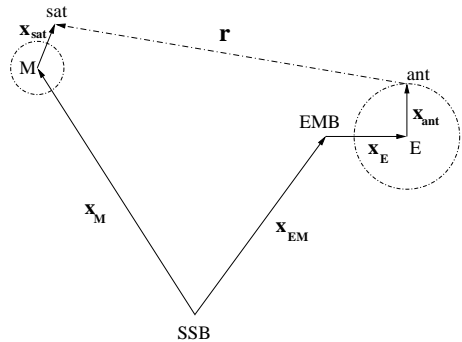}
\end{center}
\caption{\footnotesize Geometric sketch of the vectors involved in the
  computation of the range. SSB is the Solar System Barycenter, M is
  the center of Mercury, EMB is the Earth-Moon Barycenter, E is the
  center of the Earth.}
\label{fig:range}
\end{figure}

How we compute these observables?  For example, a first approximation
of the range could be given by the formula
\begin{equation}
r=|{\bf r}|=|({\bf x}_{\rm sat}+{\bf x}_{\rm M})-({\bf x}_{\rm EM}+
{\bf x}_{\rm E}+{\bf x}_{\rm ant})| \,\, ,
\label{eq:new}
\end{equation}
which models a very simple geometrical situation (see
Figure~\ref{fig:range}). The vector ${\bf x}_{\rm sat}$ is the
mercurycentric position of the orbiter, the vector ${\bf x}_{\rm M}$
is the position of the center of mass of Mercury (M) in a reference
system with origin at the Solar System Barycenter (SSB), the vector
${\bf x}_{\rm EM}$ is the position of the Earth-Moon center of mass in
the same reference system, ${\bf x}_{\rm E}$ is the vector from the
Earth-Moon Barycenter (EMB) to the center of mass of the Earth (E),
the vector ${\bf x}_{\rm ant}$ is the position of the reference point
of the ground antenna with respect to the center of mass of the Earth.

Using (\ref{eq:new}) means to model the space as flat arena ($r$ is an
Euclidean distance) and the time as absolute parameter.  This is
obviously not possible because it is clear that, beyond some threshold
of accuracy, these quantities have to be formulated within the
framework of Einstein's theory of gravity (general relativity theory,
GRT). Moreover we have to take into account the different times at
which the events have to be computed: the transmission of the signal
at the transmit time ($t_t$), the signal at the Mercury orbiter at the
time of bounce ($t_b$) and the reception of the signal at the receive
time ($t_r$).

Formula (\ref{eq:new}) could be a good starting point to construct a
correct relativistic formulation; with the word ``correct'' we do not
mean all the possible relativistic effects, but the effects that can
be measured by the experiment. This paper deals with the corrections
to apply to this formula to obtain a consistent relativistic model for
the computations of the observables and the practical implementation
of such computations.

In Section \ref{sec:refsys} we discuss the relativistic
four-dimensional reference systems used and the transformations
adopted to make the sums in (\ref{eq:new}) consistent; according to
\cite{soffel03}, with ``reference system'' we mean a purely
mathematical construction, while a ``reference frame'' is a some
physical realization of a reference system. The relativistic
contribution to the time delay due to the Sun's gravitational field,
the Shapiro effect, is described in Section \ref{sec:shap}.  Section
\ref{sec:lt} deals with the theoretical procedure to compute the
light-time (range) and the Doppler shift (range rate).  In
Section~\ref{sec:num} we discuss the practical implementation of the
algorithms showing how we eliminate rounding-off problems.

The equations of motion for the planets Mercury and Earth, including
all the relativistic effects (and potential violations of GRT)
required to the accuracy of the BepiColombo radioscience experiment
have already been discussed in \cite{mil09b}, thus this paper
concentrates on the computation of the observables.

\section{Space-time reference frames and transformations}
\label{sec:refsys}

The five vectors involved in formula (\ref{eq:new}) have to be
computed at their own time, the epoch of different events: e.g., ${\bf
x}_{\rm ant}$, ${\bf x}_{\rm EM}$ and ${\bf x}_{\rm E}$ are computed
at both the antenna \emph{transmit time} $t_t$ and the \emph{receive
time} $t_r$ of the signal. ${\bf x}_{\rm M}$ and ${\bf x}_{\rm sat}$
are computed at the \emph{bounce time} $t_b$ (when the signal has
arrived to the orbiter and is sent back, with correction for the delay
of the transponder). To be able to perform the vector sums and
differences, these vectors have to be converted to a common space-time
reference system, the only possible choice being some realization of
the BCRS (Barycentric Celestial Referece System). We adopt for now a
realization of the BCRS that we call SSB (Solar System Barycentric)
reference frame and in which the time is TDB (Barycentric Dynamic
Time); other possible choices, such as a TCB (Barycentric Celestial
Time), only can differ by linear scaling. The TDB choice of the SSB
timescale entails also the appropriate linear scaling of
space-coordinates and planetary masses as described for instance
in \cite{kli08} or \cite{kli09}.

The vectors ${\bf x}_{\rm M}$, ${\bf x}_{\rm E}$, and ${\bf x}_{\rm
EM}$ are already in SSB as provided by numerical integration and
external ephemerides; thus the vectors ${\bf x}_{\rm ant}$ and ${\bf
x}_{\rm sat}$ have to be converted to SSB from the geocentric and
mercurycentric systems, respectively. Of course the conversion of
reference system implies also the conversion of the time coordinate.
There are three different time coordinates to be considered. The
currently published planetary ephemerides are provided in TDB.  The
observations are based on averages of clock and frequency measurements
on the Earth surface: this defines another time coordinate called
TT (Terrestrial Time).  Thus for each observation the times of
transmission $t_t$ and receiving $t_r$ need to be converted from TT to
TDB to find the corresponding positions of the planets, e.g., the
Earth and the Moon, by combining information from the precomputed
ephemerides and the output of the numerical integration for Mercury
and the Earth-Moon barycenter. This time conversion step is necessary
for the accurate processing of each set of interplanetary tracking
data; the main term in the difference TT-TDB is periodic, with period
1 year and amplitude $\simeq 1.6\times 10^{-3}$ s, while there is
essentially no linear trend, as a result of a suitable definition of
the TDB.

The equation of motion of a mercurycentric orbiter can be
approximated, to the required level of accuracy, by a Newtonian
equation provided the independent variable is the proper time of
Mercury.  Thus, for the BepiColombo radioscience experiment, it is
necessary to define a new time coordinate TDM (Mercury Dynamic Time),
as described in \cite{mil09b}, containing terms of 1-PN
order depending mostly upon the distance from the Sun and velocity of
Mercury.

 From now on we shall call the quantities related to the SSB frame
``TDB-compatible'', the quantities related to the geocentric frame
``TT-compatible'', and the quantities related to the mercurycentric
frame ``TDM-compatible'', in accordance with the paper \cite{kli09},
and label them TB, TT and TM, respectively.

The differential equation giving the local time $T$ as a function of the 
SSB time $t$ , which we are currently assuming to be TDB, is the following:
\begin{equation}
\dertot{T}{t}= 1- \frac{1}{c^2}\;\left[U+ \frac{v^2}{2}-L\right]\,\,,
\label{eq:dert}
\end{equation}
where $U$ is the gravitational potential (the list of contributing
bodies depends upon the accuracy required: in our implementation we
use Sun, Mercury to Neptune, Moon) at the planet center and $v$ is the
SSB velocity of the same planet. The constant term $L$ is used to
perform the conventional rescaling motivated by removal of secular
terms, e.g., for the Earth we use $L_C$.

The space-time transformations we have to perform involve essentially
the position of the antenna and the position of the orbiter.  The
geocentric coordinates of the antenna should be transformed into
TDB-compatible coordinates; the transformation is expressed by the
formula
 \[
 {\bf x}_{\rm ant}^{TB}={\bf x}_{\rm ant}^{TT}\,\left
 (1-\frac{U}{c^2}-L_C \right)- \frac{1}{2}\,\left ( \frac{{\bf v}_{\rm
 E}^{TB}\cdot {\bf x}_{\rm ant}^{TT}}{c^2} \right )\, {\bf v}_{\rm
 E}^{TB}\,\,,
 \]
where $U$ is the gravitational potential at the geocenter (excluding
the Earth mass), $L_C=1.48082686741\, \times \,10^{-8} $ is a scaling
factor given as definition, supposed to be a good approximation for
removing secular terms from the transformation and ${\bf v}_{\rm
E}^{TB}$ is the barycentric velocity of the Earth. The next formula
contains the effect on the velocities of the time coordinate change,
which should be consistently used together with the coordinate change:
 \[
 {\bf v}_{\rm ant}^{TB} = \left[{\bf v}_{\rm ant}^{TT}\,\left
 (1-\frac{U}{c^2}-L_C \right)- \frac{1}{2}\,\left ( \frac{{\bf v}_{\rm
 E}^{TB}\cdot {\bf v}_{\rm ant}^{TT}}{c^2} \right )\, {\bf v}_{\rm
 E}^{TB}\right] \cdot \left[\dertot Tt\right]\,\,.
 \]
Note that the previous formula contains the factor $dT/dt$ (expressed
by eq.~(\ref{eq:dert})) that deals with time transformation: $T$ is
the local time for Earth, that is TT, and $t$ is the corresponding
TDB time.

The mercurycentric coordinates of the orbiter should be transformed
into TDB-compatible coordinates through the formula
 \[
 {\bf x}_{\rm sat}^{TB}={\bf x}_{\rm sat}^{TM}\,\left (1-\frac{U}{c^2}-L_{CM}
 \right)- \frac{1}{2}\,\left ( \frac{{\bf v}_{\rm M}^{TB} \cdot {\bf
 x}_{\rm sat}^{TM}}{c^2} \right )\, {\bf v}_{\rm M}^{TB}\,\,,
 \]
where $U$ is the gravitational potential at the center of mass of
Mercury (excluding the Mercury mass) and $L_{CM}$ could be used to
remove the secular term in the time transformation (thus defining a TM
scale, implying a rescaling of the mass of Mercury). We believe this
is not necessary: the secular drift of TDM with respect to other time
scales is significant, see Figure 5 in paper \cite{mil09b}, but a
simple iterative scheme is very efficient in providing the inverse
time transformation.  Thus we set $L_{CM}=0$, assuming the reference
frame is TDM-compatible.
As for the antenna we have a formula expressing the velocity
transformation that contains the derivative of time $T$ for Mercury,
that is TDM, with respect to time $t$, that is TDB:
 \[
 {\bf v}_{\rm sat}^{TB} =  
 \left[ {\bf v}_{\rm sat}^{TM}\,\left
 (1-\frac{U}{c^2}-L_{CM} \right)- \frac{1}{2}\,\left ( 
 \frac{{\bf  v}_{\rm M}^{TB}\cdot {\bf v}_{\rm sat}^{TM}}{c^2} 
 \right )\,{\bf v}_{\rm M}^{TB} \right ] \cdot
 \left[\dertot{T}{t} \right]\,\, .
 \]

In all the formulas for these coordinate changes we have neglected
the terms of the SSB acceleration of the planet center
(\cite{dam94}), because they contain beside $1/c^2$ the additional
small parameter (distance from planet center)/(planet distance to the
Sun), which is of the order of $10^{-4}$ even for a Mercury orbiter.

\begin{figure}
\begin{center}
\begin{tabular}{c}
 \includegraphics[width=7cm]{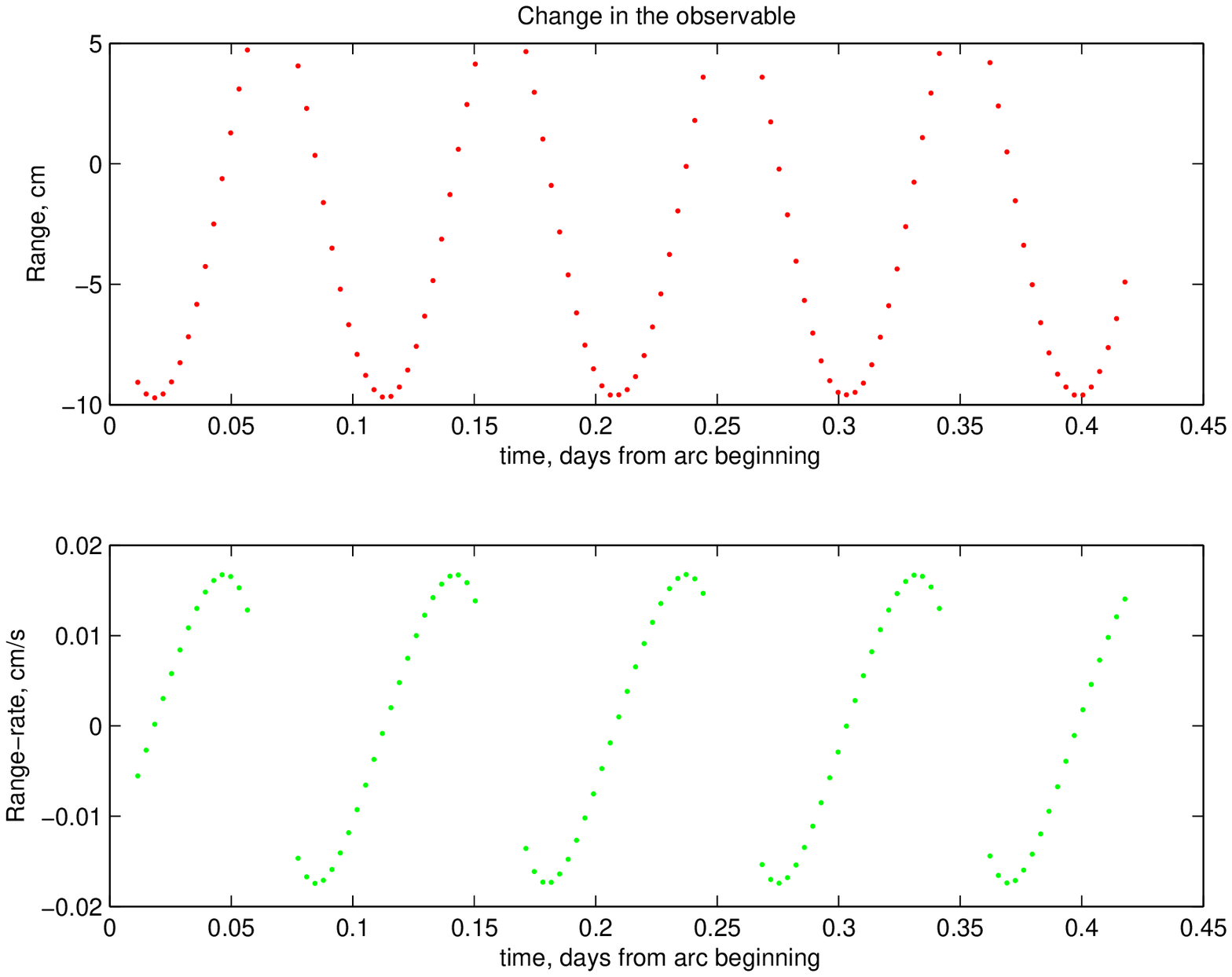} \\
 \includegraphics[width=7cm]{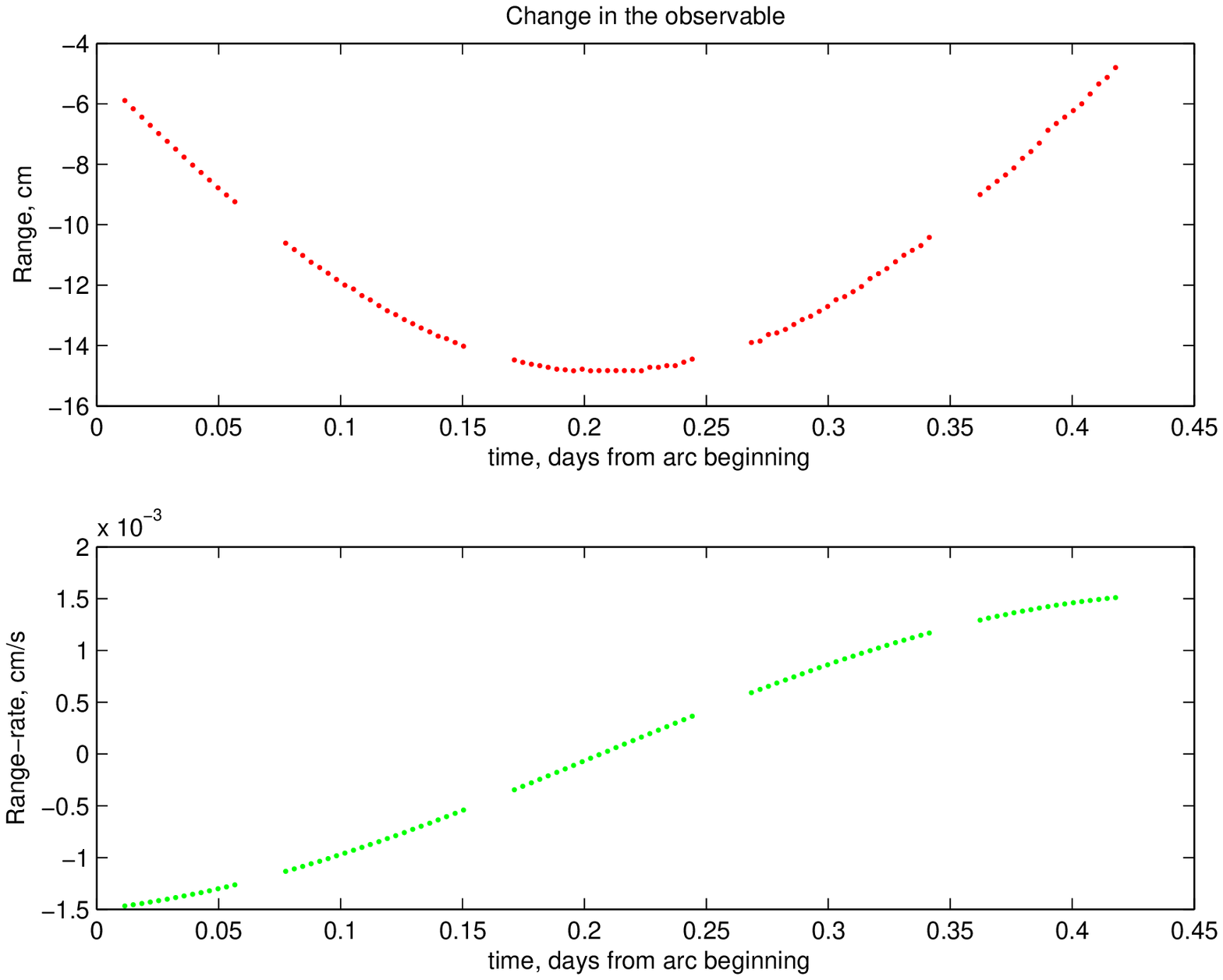} 
\end{tabular}
\end{center}
\caption{\footnotesize The difference in the observables range and
  range rate for one pass of Mercury above the horizon for a ground
  station, by using an hybrid model in which the position and velocity
  of the orbiter have not transformed to TDB-compatible quantities and
  a correct model in which all quantities are
  TDB-compatible. Interruptions of the signal are due to spacecraft
  passage behind Mercury as seen for the Earth station. Top: for an
  hybrid model with the satellite position and velocity not
  transformed to TDB-compatible. Bottom: for an hybrid model with the
  position and velocity of the antenna not transformed to
  TDB-compatible.}
\label{fig:spcoord}
\end{figure}

To assess the relevance of the relativistic corrections of this
section to the accuracy of the BepiColombo radioscience experiment, we
have computed the observables range and range rate with and without
these corrections. As shown in Figure~\ref{fig:spcoord}, the
differences are significant, at a signal-to-noise ratio $S/N\simeq 1$ for
range, much more for range rate, with an especially strong signature from
the orbital velocity of the mercurycentric orbit (with $S/N> 50$).

\section{Shapiro effect}
\label{sec:shap}

\begin{figure}
\begin{center}
\includegraphics[width=7cm]{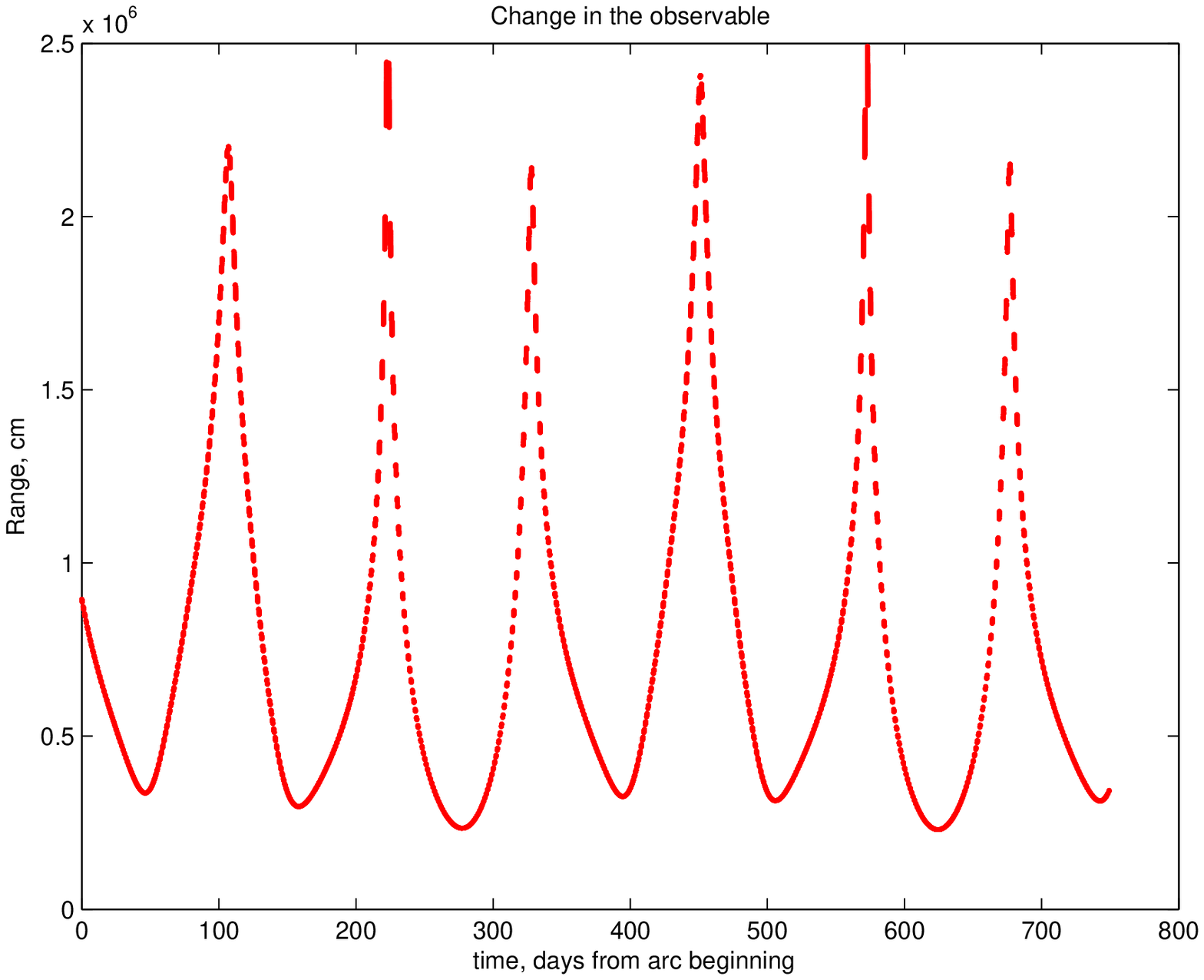}
\end{center}
\caption{\footnotesize Total amount of the Shapiro effect in range
over 2-year simulation. The sharp peaks correspond to superior
conjunctions, when Mercury is ``behind the Sun'' as seen from Earth,
with values as large as $24$ km for radiowaves passing at 3 solar
radii from the center of the Sun. Interruptions of the signal are due
to spacecraft visibility from the Earth station (in this simulation we
assume just one station).}
\label{fig:shap1pn}
\end{figure}

The correct modelling of space-time transformations is not sufficient
to have a precise computation of the signal delay: we have to take
into account the general relativistic contribution to the time delay
due to the space-time curvature under the effect of the Sun's
gravitational field, the {\em Shapiro effect} (\cite{shap64}). The
Shapiro time delay $\Delta t$ at the 1-PN level, according to
\cite{will93} and \cite{moyer03}, is
\[
\Delta t=\frac{(1+\gamma)\,\mu_0}{c^3}\, \ln \, \left (
\frac{r_t+r_r+r}{r_t+r_r-r} \right )\,, \quad 
S(\gamma)=c\,\Delta t \,\,;
\]
$r_t=|{\bf r}_{\rm t}|$ and $r_r=|{\bf r}_{\rm r}|$ are the
heliocentric distances of the transmitter and the receiver at the
corresponding time instants of photon transmission and reception,
$\mu_0$ is the gravitational mass of the Sun ($\mu_0=G\,m_0$) and
$r=|{\bf r}_r-{\bf r}_t|$.  The planetary terms, similar to the solar
one, can also be included but they are smaller than the accuracy
needed for our measurements. Parameter $\gamma$ is the only
post-Newtonian parameter used for the light-time effect and, in fact,
it could be best constraint during superior conjunction
(\cite{mil02}). The total amount of the Shapiro effect in range is
shown in Figure \ref{fig:shap1pn}.

The question arises whether the very high signal to noise in the range
requires other terms in the solar gravity influence, due to either (i)
motion of the source, or (ii) higher-order corrections when the radio
waves are passing near the Sun, at just a few solar radii (and thus
the denominator in the log-function of the Shapiro formula is
small). The corrections (i) are of the post-Newtonian order 1.5, that
is containing a factor $1/c^3$, but it has been shown in \cite{mil09b}
that they are too small to affect our accuracy. The corrections (ii)
are of order 2, that is $1/c^4$, but they can be actually larger for
an experiment involving Mercury. The relevant correction is most
easily obtained by adding $1/c^4$ terms in the Shapiro formula, due to
the bending of the light path:
\[
S(\gamma)=\frac{(1+\gamma)\,\mu_0}{c^2}\, \ln
\,\left(\frac{r_t+r_r+r+\frac{(1+\gamma)\,\mu_0}{c^2}}
{r_t+r_r-r+\frac{(1+\gamma)\,\mu_0}{c^2}}\right) \,\,.
\]

\begin{figure}[h]
\begin{center}
\begin{tabular}{c}
\includegraphics[width=7cm]{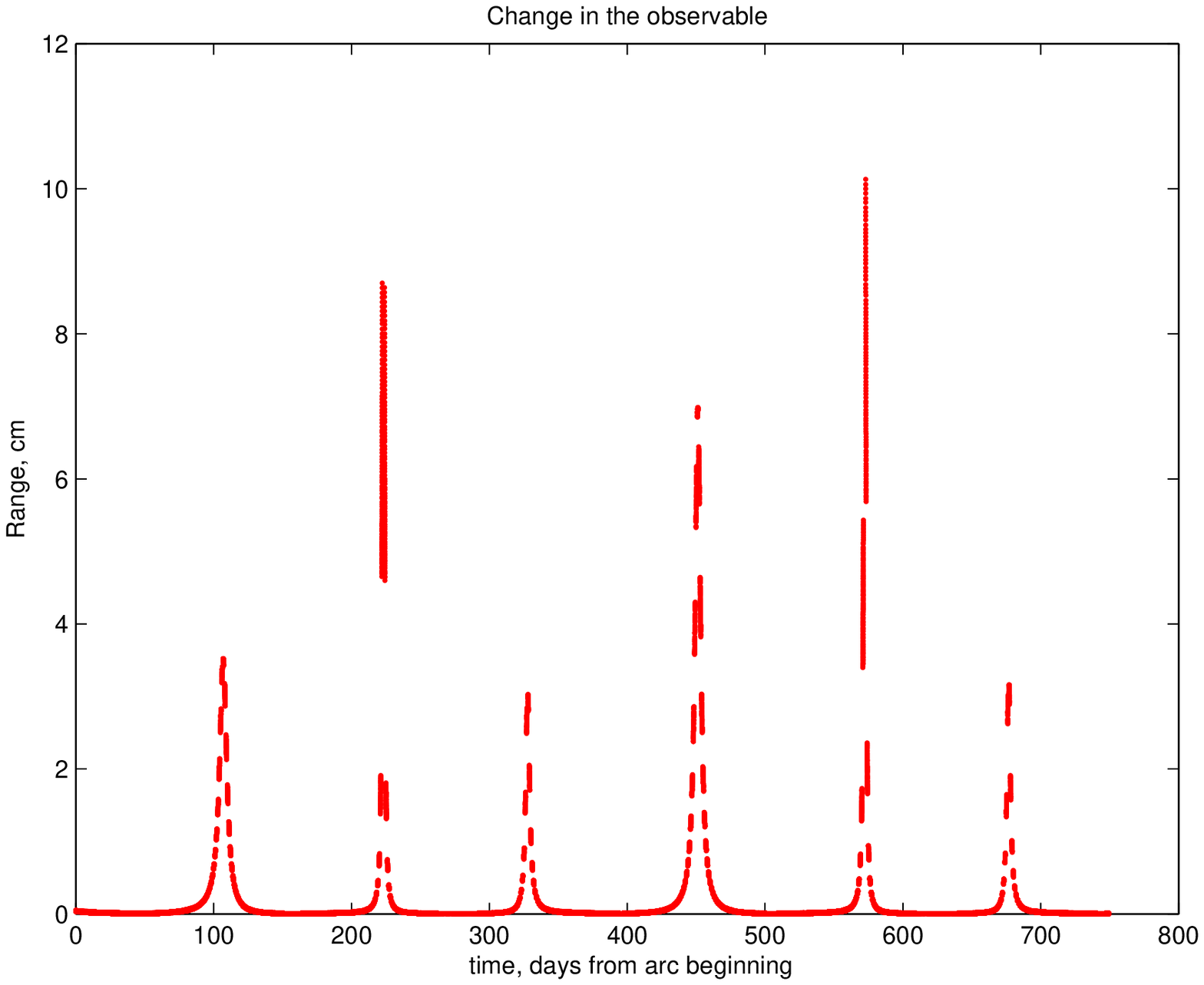}\\
\includegraphics[width=7cm]{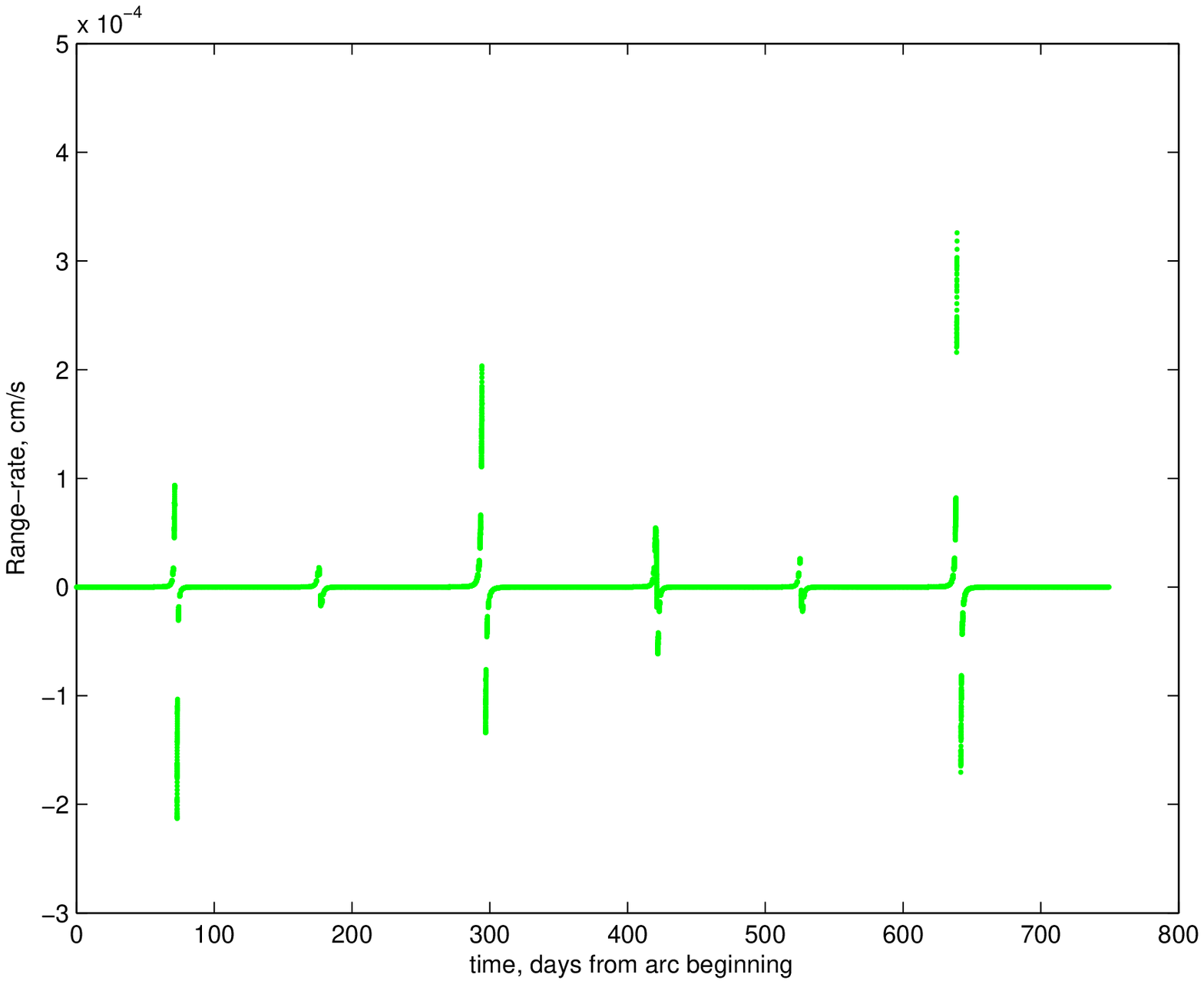}
\end{tabular}
\end{center}
\caption{\footnotesize Differences in range (top) and range rate
  (bottom) by using an order 1 and an order 2 post-Newtonian
  formulation ($\gamma=1$); the correction is relevant for
  BepiColombo, at least when a superior conjunction results in a small
  impact parameter $b$. E.g., in this figure we have plotted data
  assumed to be available down to $b\simeq 3 R_0$. For larger values
  of $b$ the effect decreases as $1/b^2$. }
\label{fig:shap}
\end{figure}

This formulation has been proposed in \cite{moyer03} and it has been
justified in the small impact parameter regime by much more
theoretically rooted derivations in \cite{klzs07}, \cite{teylepon08}
and \cite{ashby08}. Figure~\ref{fig:shap} shows that the order 2
correction is relevant for our experiment, especially when there is a
superior conjunction with a small impact parameter of the radio wave
path passing near the Sun. Note that the $1/c^4$ correction ($\sim
10$~cm) in the Shapiro formula effectively corresponds to $\sim
3\times 10^{-5}$ correction in the value of the post-Newtonian
parameter $\gamma$.

The Shapiro correction for the computation of the range rate is
\[
\dot S=\frac{2(1+\gamma)\mu_0}{c^2}\left[ \frac{-r\,(\dot r_t+\dot
r_r)+\dot r\,\left(r_t+r_r+
\frac{(1+\gamma)\mu_0}{c^2}\right)}{(r_t+r_r+\frac{(1+\gamma)\mu_0}
{c^2})^2-r^2}
\right] \,\,,
\]
where the time derivative is with respect to a TDB time. This formula
is almost never found in the literature and has not been much used in
the processing of the past radioscience experiments, such as in
\cite{bertotti03}, because the observable range rate is typically
computed as difference of ranges divided by time; however, for reasons
explained in Section~\ref{sec:num}, this formula is now necessary.

\section{Light-time iterations}
\label{sec:lt}


Since radar measurements are usually referred to the receive time
$t_r$ the observables are seen as functions of this time, and the
computation sequence works backward in time: starting from $t_r$, the
bounce time $t_b$ is computed iteratively, and, using this information
the transmit time $t_t$ is computed. 

The vectors ${\bf x}_{\rm M}^{TB}$ and ${\bf x}_{\rm EM}^{TB}$ are
obtained integrating the post-Newtonian equations of motions. The
vectors ${\bf x}_{\rm sat}^{TM}$ are obtained by integrating the orbit
in the mercurycentric TDM-compatible frame. The vector ${\bf x}_{\rm
ant}^{TT}$ is obtained from a standard IERS model of Earth rotation,
given accurate station coordinates, and ${\bf x}_{\rm E}^{TT}$ from
lunar ephemerides (\cite{mil09a}).

In the following subsections we shall describe the procedure to
compute the range (Section \ref{sec:range}) and the range rate
(Section \ref{sec:ranger}).

\subsection{Range}
\label{sec:range}

\begin{figure}
\begin{center}
\includegraphics[width=7cm]{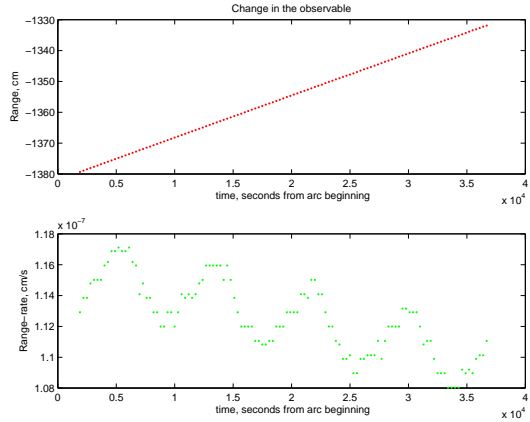}
\end{center}
\caption{\footnotesize The difference in the observables range and
  range rate using a light-time in TT and a light time in TDB: the
  difference in range is very high, more than 13 meters in one day,
  while the difference in range rate is less than the accuracy of the
  experiment.}
\label{fig:difft}
\end{figure}

Once the five vectors are available at the appropriate times and in a
consistent SSB system, there are two different light-times, the up-leg
$\Delta t_{up}= t_b-t_t$ for the signal from the antenna to the
orbiter, and the down-leg $\Delta t_{down}= t_r-t_b$ for the return
signal. They are defined implicitly by the distances down-leg and
up-leg
\begin{eqnarray}
{\bf r}_{do}(t_r)&=& {\bf x}_{\rm sat}(t_b(t_r))+{\bf x}_{\rm
M}(t_b(t_r))- {\bf x}_{\rm EM}(t_r)-{\bf x}_{\rm E}(t_r)-{\bf x}_{\rm
ant}(t_r) \,\, ,\nonumber\\ r_{do}(t_r)&=&|{\bf r}_{do}(t_r)|\,\, ,
\quad \quad c(t_r-t_b)=r_{do}(t_r) +S_{do}(\gamma)\,\, ,
\label{eq:rdo}
\end{eqnarray}
\begin{eqnarray}
{\bf r}_{up}(t_r)&=& {\bf x}_{\rm sat}(t_b(t_r))+{\bf x}_{\rm
M}(t_b(t_r))- {\bf x}_{\rm EM}(t_t(t_r))-{\bf x}_{\rm E}(t_t(t_r))-{\bf
x}_{\rm ant}(t_t(t_r))\,\, , \nonumber\\ r_{up}(t_r)&=&|{\bf r}_{up}(t_r)
|\,\, , \quad \quad c(t_b-t_t)=r_{up}(t_r) +S_{up}(\gamma)\,\, ,
\label{eq:rup}
\end{eqnarray}
respectively, with somewhat different Shapiro effects. Then $t_r-t_b$
and $t_b-t_t$ are the two portions of the light-time, in the time
attached to the SSB, that is TDB; this provides the computation of
$t_t$. Then these times are to be converted back in the time system
applicable at the receiving station, where the time measurement is
performed, which is TT (or some other form of local time, such as the
standard UTC). $t_r$ is already available in the local time scale,
from the original measurement, while $t_t$ needs to be converted back
from TDB to TT. The difference between these two TT times is $\Delta
t_{tot}$, from which we can conventionally define $r(t_r)=c\, \Delta
t_{tot}/2$. Note that the difference $\Delta t_{tot}$ in TT is
significantly different from $t_r-t_t$ in TDB, by an amount of the
order of $10^{-7}$ s, while the sensitivity of the BepiColombo
radioscience experiment is of the order of $10^{-9}$ s, thus these
conversions change the computed observable in a significant way, see
Figure~\ref{fig:difft}.

The practical method for solving $t_b(t_r)$ and $t_t(t_r)$ in
eqs.~(\ref{eq:rdo}) and (\ref{eq:rup}) is as follows. Since the
measurement is labeled with the receive time $t_r$, the iterative
procedure needs to start from eq.~(\ref{eq:rdo}) by computing the
states ${\bf x}_{\rm EM},\;{\bf x}_{\rm E}$ and ${\bf x}_{\rm ant}$ at
epoch $t_r$, then selecting a rough guess $t_b^0$ for the bounce time
(e.g., $t^0_b=t_r$).  Then the states ${\bf x}_{\rm sat}$ and ${\bf
x}_{\rm M}$ are computed at $t_b^0$ and a successive guess $t_b^1$ is
given by (\ref{eq:rdo}). This is repeated computing $t_b^2$, and so on
until convergence, that is, until $t_b^k-t_b^{k-1}$ is smaller than
the required accuracy. This fixed point iteration to solve the
implicit equation for $t_b$ is convergent because the motion of the
satellite and of Mercury, in the time $t_r-t_b$, is a small fraction
of the total difference vector.  After accepting the last value of
$t_b$ we start with the states ${\bf x}_{\rm sat}$ and ${\bf x}_{\rm
M}$ at $t_b$ and with a rough guess $t_t^0$ for the transmit time
(e.g., $t^0_t=t_b$).  Then ${\bf x}_{\rm EM},\;{\bf x}_{\rm E}$ and
${\bf x}_{\rm ant}$ are computed at epoch $t_t^0$ and $t_t^1$ is given
by eq.~(\ref{eq:rup}), and the same procedure is iterated to
convergence, that is to achieve a small enough $t_t^k-t_t^{k-1}$. This
double iterative procedure to compute range is consistent with what
has been used for a long time in planetary radar, as described in the
paper \cite{yeo92}. We conventionally define $r
=(r_{do}+S_{do}+r_{up}+ S_{up})/2$.

\subsection{Range rate}
\label{sec:ranger}

After the two iterations providing at convergence $t_b$ and $t_t$ are
complete, we can proceed to compute the range rate. We rewrite the
expression for the Euclidean range (down-leg and up-leg) as a scalar product:
\begin{eqnarray*}
r^2_{do}(t_r)&=&\left [ {\bf x}_{\rm Ms}(t_b) -{\bf x}_{\rm Ea}(t_r)
  \right ] \, \cdot \, \left [{\bf x}_{\rm Ms}(t_b) -{\bf x}_{\rm
    Ea}(t_r) \right] \,\, ,\\ r^2_{up}(t_r) & = & \left [ {\bf x}_{\rm
    Ms}(t_b)-{\bf x}_{\rm Ea}(t_t)\right] \cdot \left [{\bf x}_{\rm
    Ms}(t_b) -{\bf x}_{\rm Ea}(t_t)\right ] \,\, ,
\end{eqnarray*}
where ${\bf x}_{\rm Ms}={\bf x}_{\rm M}+{\bf x}_{\rm sat}$ and ${\bf
  x}_{\rm Ea}={\bf x}_{\rm EM}+{\bf x}_{\rm E}+{\bf x}_{\rm ant}$.
  The light-time equation contains also the Shapiro terms, thus the
  range rate observable contains also additive terms $\dot S_{do}$ and
  $\dot S_{up}$, with significant effects (a few cm/s during superior
  conjunctions).  Since the equations giving $t_b$ and $t_t$ are still
  (\ref{eq:rdo}) and (\ref{eq:rup}), in computing the time
  derivatives, we need to take into account that $t_b=t_b(t_r)$ and
  $t_t=t_t(t_r)$, with non-unit derivatives.

Computing the derivative with respect to the receive time $t_r$, and
using the dot notation to stand for $d/dt_r$, we obtain:
 \begin{eqnarray}\label{eq:rdodot}
 \dot r_{do}(t_r)& =& \hat {\bf r}_{do}\, \left[ {\bf \dot x}_{\rm
 Ms}(t_b)\, \left (1-\frac{\dot r_{do}(t_r)+\dot S_{do}}{c} \right )-
 {\bf \dot x}_{\rm Ea}(t_r)\right] \,\, , \\
  \dot r_{up}(t_r)& = &\hat {\bf r}_{up}\,\Bigg [ {\bf \dot x}_{\rm
 Ms}(t_b)\, \left (1-\frac{\dot r_{do}(t_r)+\dot S_{do}}{c} \right )-
 \nonumber
\end{eqnarray}
\begin{equation}\label{eq:rupdot}
 \quad \quad \quad \quad \quad{\bf \dot x}_{\rm Ea}(t_t) \left (1-\frac{\dot
 r_{do}(t_r)+ \dot S_{do}}{c}-\frac{\dot r_{up}(t_r)+ \dot S_{up}}{c}
 \right ) \Bigg ]\,\, ,
 \end{equation}
where 
\[
\hat {\bf r}_{do}=\frac{{\bf x}_{\rm Ms}(t_b)-{\bf x}_{\rm
Ea}(t_r)}{r_{do}(t_r)} \,\, , \quad \quad
\hat {\bf r}_{up}=\frac{{\bf x}_{\rm Ms}(t_b)-{\bf x}_{\rm
Ea}(t_t)}{r_{up}(t_r)} \,\, .
\]

\noindent However, the contribution of the time derivatives of the
Shapiro effect to the $d\,t_b/d\,t_r$ and $d\,t_t/d\,t_r$ corrective
factors is small, of the order of $10^{-10}$, which is marginally
significant for the BepiColombo radioscience experiment. We
conventionally define $\dot r =c(1-\dot t_t)/2=(\dot r_{do}+\dot
S_{do}+\dot r_{up}+\dot S_{up})/2$.  These equations are compatible
with \cite{yeo92}, taking into account that they use a single
iteration.

Since the time derivatives of the Shapiro effects contain $\dot r_t$
and $\dot r_r$, the equations (\ref{eq:rdodot}) and (\ref{eq:rupdot})
are implicit, thus we can again use a fixed point iteration. It is
also possible to use a very good approximation which solves explicitly
for $\dot r_{do}$ and then for $\dot r_{up}$, neglecting the very
small contribution of Shapiro terms:
\[
\dot r_{do}= \hat{\bf r}_{do}\cdot\left[{\bf \dot
x}_{\rm Ms}(t_b)\,\left(1-\frac{\dot S_{do}}{c}\right)-{\bf \dot
x}_{\rm Ea}(t_r)\right]\;\left[1 +\frac{{\bf \dot x}_{\rm Ms}(t_b)\cdot
\hat{\bf r}_{do}}{c} \right]^{-1} \,\, ,
\]
where the right hand side is weakly dependent upon $\dot r_{do}$ only
through $\dot S_{do}$, thus a moderately accurate approximation could
be used in the computation of $\dot S_{do}$, followed by a single
iteration. For the other leg
 \begin{eqnarray*}
 \dot r_{up}(t_r) &=&  
 \hat{\bf r}_{up} \cdot
 \left [ {\bf \dot x}_{\rm Ms}(t_b)\, \left (1-\frac{\dot r_{do}(t_r)+\dot
 S_{do}}{c} \right ) -{\bf \dot x}_{\rm Ea}(t_t) \left (1-\frac{\dot
 r_{do}(t_r)+ \dot S_{do}}{c}-\frac{\dot S_{up}}{c}
 \right ) \right ] \\
 && \quad \quad \, \left[
 1-\frac{{\bf \dot x}_{\rm Ea}(t_t) \cdot\hat{\bf r}_{up}}{c}
 \right]^{-1} \,\,.
 \end{eqnarray*}

All the above computations are in SSB with TDB; however, the frequency
measurements, at both $t_t$ and $t_r$, are done on Earth, that is with
a time which is TT. This introduces a change in the measured
frequencies at both ends, and because this change is not the same (the
Earth having moved by about $3\times 10^{-4}$ of its orbit) there is a
correction needed to be performed. The quantity we are measuring is
essentially the derivative of $t_t$ with respect to $t_r$, but this in
two different time systems: for readability, we use $T$ for TT, $t$
for TDB
\[
\dertot{T_t}{T_r}=
\dertot{T_t}{t_t}\;\dertot{t_t}{t_r}\;\dertot{t_r}{T_r}\,\,,
\]
where the derivatives of the time coordinate changes are the same as
the right hand sides of the differential equation giving $T$ as a
function of $t$ in the first factor and the inverse of the same for
the last factor. However, the accuracy required is such that the main
term with the gravitational mass of the Sun $\mu_0$ and the position
of the Sun ${\bf x}_0$ is enough:
 \begin{eqnarray}
 \dertot{T_t}{T_r} & = &
 \left[1-\frac{\mu_0}{|{\bf x}_E(t_t)-{\bf x}_0(t_t)|\,c^2}-
 \frac{|{\bf \dot x}_{\rm E}(t_t)|^2}{2\,c^2}\right] 
 \;\dertot{t_t}{t_r} \nonumber \\
 && \left[1-\frac{\mu_0}{|{\bf x}_{\rm E}(t_r)-{\bf x}_0(t_r)|\,c^2}-
 \frac{|{\bf \dot x}_{\rm E}(t_r)|^2}{2\,c^2}\right]^{-1}\ .
 \label{eq:dertimeconv}
 \end{eqnarray}
Note that we do not need the $L_C$ constant term discussed above
because it cancels in the first and last terms in the right hand sides
of eq.~(\ref{eq:dertimeconv}). The correction in the above formula is
required for consistency, but in fact the correction has an order of
magnitude of $10^{-7}$ cm/s and is negligible for the sensitivity of
the BepiColombo radioscience experiment (Figure~\ref{fig:difft}).

\section{Numerical problems and solutions}
\label{sec:num}

The computation of the observables, as presented in the previous
section, is already complex, but still the list of subtle
technicalities is not complete.  

A problem well known in radioscience is that, for top accuracy, the
range rate measurement cannot be the instantaneous value $\dot
r(t_r)=(\dot r_{do}(t_r)+\dot S_{do}+\dot r_{up}(t_r)+\dot S_{up})/2$.
In fact, the measurement is not instantaneous: an accurate measure of
a Doppler effect requires to fit the difference of phase between
carrier waves, the one generated at the station and the one returned
from space, accumulated over some \emph{integration time} $\Delta$,
typically between $10$ and $1000$ s. Thus the observable is really a
difference of ranges
\begin{equation}
\dot r_{\Delta}(t_r)=\frac{r(t_b+\Delta/2)-r(t_b-\Delta/2)}{\Delta}
\label{eq:rrate_mode3}
\end{equation}
or, equivalently, an averaged value of range rate over the integration interval
\begin{equation}
\dot r_{\Delta}(t_r)=\frac 1\Delta\; \int_{t_b-\Delta/2}^{t_b+\Delta/2}\;
\dot r(s)\; ds\ .
\label{eq:rrate_mode4}
\end{equation}

\begin{figure}[h]
\begin{center}
\includegraphics[width=7cm]{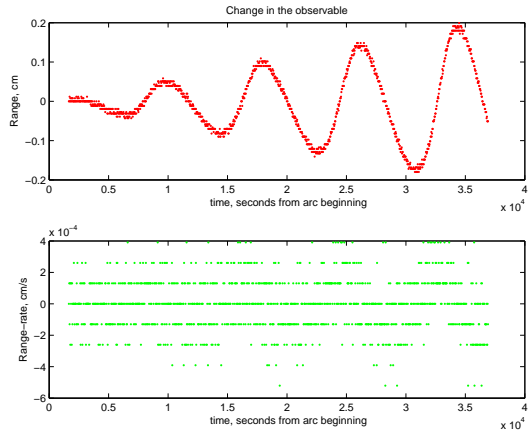} 
\end{center}
\caption{\footnotesize Range and range rate differences due to a
change by $10^{-11}$ of the $C_{22}$ harmonic coefficient: the range
rate computed as range difference divided by the integration time of
30 s, eq.~(\ref{eq:rrate_mode3}), is obscured by the rounding off.}
\label{fig:rr3}
\end{figure}

In order to understand the computational difficulty we need to take
also into account the orders of magnitude. As said in the
introduction, for state of the art tracking systems, such as those
using a multi-frequency link in the X and Ka bands, the accuracy of
the range measurements can be $\simeq 10$ cm and the one of range-rate
$3\times 10^{-4}$ cm/s (over an integration time of $1\,000$ s).  Let
us take an integration time $\Delta=30$ s, which is adequate for
measuring the gravity field of Mercury; in fact if the orbital period
is $\simeq 8\,000$ s, the harmonics of order $m=26$ have periods as
short as $\simeq 150$ s.  The accuracy over $30$ s of the range rate
measurement can be, by Gaussian statistics, $\simeq 3\times
10^{-4}\;\sqrt{1\,000/30}\simeq 17 \times 10^{-4}$ cm/s, and the
required accuracy in the computation of the difference
$r(t_b+\Delta/2)-r(t_b-\Delta/2)$ is $\simeq 0.05$ cm. The distances
can be as large as $\simeq 2\times 10^{13}$ cm, thus the relative
accuracy in the difference needs to be $2.5\times 10^{-15}$. This
implies that \emph{rounding off} is a problem with current computers,
with relative rounding off error of $\varepsilon=2^{-52}=2.2\times
10^{-16}$ (Figure~\ref{fig:rr3}); extended precision is supported in
software, but it has many limitations.  The practical consequences are
that the computer program processing the tracking observables, at this
level of precision and over interplanetary distances, needs to be a
mixture of ordinary and extended precision variables.  Any
imperfection may result in ``banding'', that is residuals showing a
discrete set of values, implying that some information corresponding
to the real accuracy of the measurements has been lost in the digital
processing.

\begin{figure}[h]
\begin{center}
\includegraphics[width=7cm]{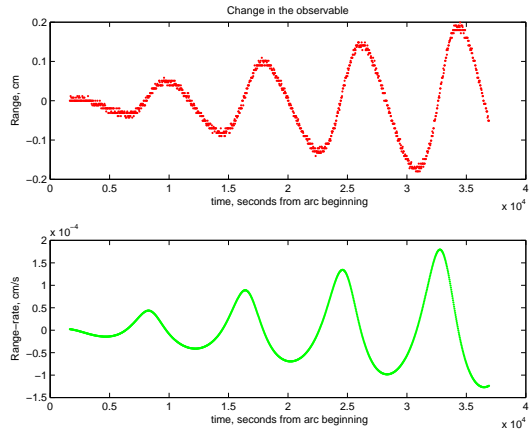}
\end{center}
\caption{\footnotesize Range and range rate differences due to a
change by $10^{-11}$ of the $C_{22}$ harmonic coefficient: the range
rate computed as an integral, eq.~(\ref{eq:rrate_mode4}), is smooth;
the difference is marginally significant with respect to the
measurement accuracy.}
\label{fig:rr4}
\end{figure}

As an alternative, the use of a quadrature formula for the integral in
eq.~(\ref{eq:rrate_mode4}) can provide a numerically more stable
result, because the S/N of the range rate measurement is
$\ll 1/\varepsilon$. Figure~\ref{fig:rr4} shows that a very small
model change, generating a range rate signal $\leq 2$ micron/s over
one pass, can be computed smoothly by using a 7 nodes Gauss quadrature
formula. 

\section{Conclusions}
\label{sec:conc}

By combining the results of the previous paper (\cite{mil09b}), and of
this one, we have completed the task of showing that it is possible to
build a consistent relativistic model of the dynamics and of the
observations for a Mercury orbiter tracked from the Earth, at a level
of accuracy and self-consistency compatible with the very demanding
requirements of the BepiColombo radioscience experiment.
 
In particular, in this paper we have given the algorithm definitions
for the computation of the observables range and range rate, including
the reference system effects and the Shapiro effect. We have shown
which computation can be performed explicitly and which ones need to
be obtained from an iterative procedure. We have also shown how to
push these computations, when implemented in a realistic computer with
rounding-off, to the needed accuracy level, even without the
cumbersome usage of quadruple precision. The list of ``relativistic
corrections'', assuming we can distinguish their effects separately,
is long, and we have shown that many subtle effects are relevant to
the required accuracy. However, in the end what is required is just to
be fully consistent with a post-Newtonian formulation to some order,
to be adjusted when necessary. Interestingly, the high accuracy of
BepiColombo radio system may require implementation of the second
post-Newtonian effects in range.


\begin{acknowledgements}
The results of the research presented in this paper, as well as in the
previous one (\cite{mil09b}) have been performed within the scope of
the contract ASI/2007/I/082/06/0 with the Italian Space
Agency. BepiColombo is a scientific space mission of the Science
Directorate of the European Space Agency. The work of DV was partially
supported by the Czech Grant Agency (grant 202/09/0772) and the
Research Program MSM0021620860 of the Czech Ministry of Education.
\end{acknowledgements}



\begin{thebibliography}{}
%
\bibitem{ashby08} Ashby, N., Bertotti, B.:
Second-order Corrections to Time Delay and Deflection of Light
Passing near a Massive Object. American Astronomical
Society, DDA meeting \#39 (2008)
%
\bibitem{bertotti03} Bertotti, B., Iess,
        L., Tortora, P.: A test of general relativity using radio
        links with the Cassini spacecraft. Nature, 425, 374-376 (2003)
%
\bibitem{dam94} Damour, T., Soffel, M., Hu,
C.: General-relativistic celestial mechanics. IV. Theory of satellite
motion. Phys. Rev. D, 49, 618-635 (1994)
%
\bibitem{iess01} Iess, L., Boscagli, G.:
Advanced radio science instrumentation for the mission BepiColombo to
Mercury. Plan. Sp. Sci., 49, 1597-1608 (2001)
%
\bibitem{klzs07} Klioner, S.A.,
  Zschocke, S.: GAIA-CA-TN-LO-SK-002-1 report (2007)
%
\bibitem{kli08} Klioner, S.A.: Relativistic scaling of
           astronomical quantities and the system of astronomical
           units. Astron. Astrophys. 478, 951--958 (2008)
%
\bibitem{kli09} Klioner, S.A., Capitaine,
	N., Folkner, W., Guinot, B., Huang, T. Y., Kopeikin, S.,
	Petit, G., Pitjeva, E., Seidelmann, P. K., Soffel, M.: Units
	of Relativistic Time Scales and Associated Quantities. In:
	Klioner, S., Seidelmann, P.K., Soffel, M. (eds.) Relativity in
	Fundamental Astronomy: Dynamics, Reference Frames, and Data
	Analysis, Cambridge University Press, in press
%
\bibitem{mil02} Milani, A.,
Vokrouhlick\'{y}, D., Villani, D., Bonanno, C., Rossi, A.: Testing
general relativity with the BepiColombo radio science
experiment. Phys. Rev. D, 66 (2002)
%
\bibitem{mil09a} Milani A., Gronchi G.F.:
Theory of orbit determination. Cambridge University Press (2009)
%
\bibitem{mil09b} Milani, A., Tommei, G.,
  Vokrouhlick\'{y}, D., Latorre, E., Cical\`o, S.: Relativistic models
  for the BepiColombo radioscience experiment. In: Klioner, S.,
  Seidelmann, P.K., Soffel, M. (eds.) Relativity in Fundamental
  Astronomy: Dynamics, Reference Frames, and Data Analysis, Cambridge
  University Press, in press
%
\bibitem{moyer03} Moyer, T.D.:
  Formulation for Observed and Computed Values of Deep Space Network
  Data Types for Navigation. Wiley-Interscience (2003)
%
\bibitem{shap64} Shapiro, I.I.: Fourth test of
general relativity. Phys. Rev. Lett., 13, 789-791 (1964)
%
\bibitem{soffel03} Soffel, M., Klioner,
  S.A., Petit, G., Kopeikin, S.M., Bretagnon, P., Brumberg, V.A.,
  Capitaine, N., Damour, T., Fukushima, T., Guinot, B., Huang, T.-Y.,
  Lindegren, L., Ma, C., Nordtvedt, K., Ries, J.C., Seidelmann, P.K.,
  Vokrouhlick\'{y}, D., Will, C.M., Xu, C.: The IAU 200 resolutions
  for astrometry, celestial mechanics, and metrology in the
  relativistic framework: explanatory supplement. Astron. J., 126,
  2687-2706 (2003)
%
\bibitem{teylepon08}
Teyssandier, P., Le Poncin-Lafitte, C.: General post-Minkowskian
expansion of time transfer functions. Class. Quantum Grav., 25, 145020
(2008)
%
\bibitem{will93} Will, C.M.: Theory and
  experiment in gravitational physics. Cambridge
  University Press (1993)
%
\bibitem{yeo92} Yeomans, D. K., Chodas,
	P. W., Keesey, M. S., Ostro, S. J., Chandler, J. F., Shapiro,
	I. I.: Asteroid and comet orbits using radar data. Astron. J.,
	103, 303-317 (1992)
\end{thebibliography}
\end{document}